# A Nanoscale Shape Memory Oxide

Jinxing Zhang[1,2*†], Xiaoxing Ke[3*†], Gaoyang Gou[4], Jan Seidel[2,5], Bin Xiang[6,9], Pu Yu[2,7], Wen-I Liang[8], Andrew M. Minor[9,10], Ying-hao Chu[8], Gustaaf Van Tendeloo[3], Xiaobing Ren[4] and Ramamoorthy Ramesh[2,10]

[1] Department of Physics, Beijing Normal University, Beijing, 100875, China

[2] Department of Physics, University of California, Berkeley, 94720, USA

[3] EMAT (Electron Microscopy for Materials Science), University of Antwerp, Groenenborgerlaan 171, Antwerp, 2020, Belgium

[4] Multi-disciplinary Materials Research Center, Frontier Institute of Science & Technology, Xi' an Jiaotong University, Xi' an 710049, China

[5] School of Materials Science and Engineering, University of New South Wales, Sydney NSW 2052, Australia

[6] CAS Key Laboratory of Materials for Energy Conversion, Department of Material Science and Engineering, University of Science and Technology of China
Hefei, Anhui, 230026, P. R. China

[7] State Key Laboratory for Low-Dimensional Quantum Physics, Department of Physics, Tsinghua University, Beijing 100084, the People's Republic of China

[8] Department of Materials Science and Engineering, National Chiao Tung University, HsinChu 30010, Taiwan

[9] National Center for Electron Microscopy, Lawrence Berkeley National Laboratory, Berkeley, 94720, USA

[10] Department of Materials Science and Engineering, University of California, Berkeley, 94720, USA

[†] These authors contributed equally to this work

[*] Email: jxzhang@bnu.edu.cn or xiaoxing.ke@uantwerpen.be

**Stimulus-responsive shape memory materials have attracted tremendous research interests recently, with much effort focused on improving their mechanical actuation. Driven by the needs of nanoelectromechnical devices, materials with large mechanical strain particularly at nanoscale are therefore desired. Here we report on the discovery of a large shape memory effect in $BiFeO_3$ at the nanoscale. A maximum strain of up to ~14% and a large volumetric work density of ~600±90 $J/cm^3$ can be achieved in association with a martensitic-like phase transformation. With a single step, control of the phase transformation by thermal activation or electric field has been reversibly achieved without the assistance of external recovery stress. Although aspects such as hysteresis, micro-cracking etc. have to be taken into consideration for real devices, the large shape memory effect in this oxide surpasses most alloys and therefore demonstrates itself as an extraordinary material for potential use in state-of-art nano-systems.**

Shape memory alloys (SMAs) offer a giant memorized mechanical response with a cycling strain of 0.5-8% [1-2]. The martensitic phase transformation, ascribed to the interplay of thermal, stress and/or magnetic fields, is responsible for the giant strain and shape memory [3]. SMAs have an indispensible role in various applications ranging from damping, mechanical joints, bio-engineering to medical applications, etc. [4-6]. In order to fulfill those functionalities, one of the most important goals that have been pursued for decades is the mechanical response under the external stimuli (electric field, magnetic field, temperature and stress) [7]. Apart from the popular alloys such as NiTi, one of the important and extensively studied model systems is $Ni_2MnGa$, where a magnetic-field-induced re-orientation of the martensitic twin walls can yield a strain as large as ~8% [8-10]. However, alloy materials exhibiting a martensitic transformation by a two-step control such as temperature/stress usually suffer from numerous problems such as surface effects, oxidation, and other instabilities, especially at the nanoscale [11-16]. One possible solution would be the exploration of the large shape memory effect (SME) /mechanical strain in an oxide material in order to enable integration with microelectronics. Microstructural analysis and theoretical studies predict that a martensitic behavior may exist in functional oxides such as manganites [17-19] and piezoelectrics [20]. Piezoelectric oxides can usually exhibit a maximum electric field induced strain of ~1.7% [21, 22] with a large susceptibility [23-24] and frequency bandwidth of their mechanical response [25-27]. Taking advantage of the rich phase diagram in conjunction with heteroepitaxial constraints, $BiFeO_3$ (BFO) can be stabilized in multiple crystal phases [28-33], namely Rhombohedral (R), Tetragonal (T) or orthorhombic (O). Especially, the coexistence of the R/T mixture indicates the possible tailoring of its phase structures [34]. However, the existence of the phase boundaries or domain walls restricts the large strain value associated with the transformation between pure phases or orientations, which is also a common problem in other ferroics or relaxors. Although the transformation from pure rhombohedral phase to pure tetragonal phase in BFO was theoretically predicted to be possible, it is still technically ambiguous due to the required ultrahigh electric field (>20 MV/cm) [35], where ferroelectric oxides under such a high electric field will suffer from dielectric breakdown. Therefore, a natural question arises: can we find a way to realize the reversible, martensitic-like transformation between those pure polymorphs with a large SME, leading to a single-step control of the large recovery strain (>10%)?

Here we report that a martensitic-like phase transformation can be achieved in a ferroelectric oxide at the nanoscale. By reducing its degree of clamping from the substrate, pure R-phase within the R/T mixture of BFO has been stabilized as observed by size-dependent atomic force microscopy (AFM) images and micro-Raman spectra. Transmission electron microscopy (TEM) provides direct evidence for this phase transformation, demonstrating a SMA-like effect in this oxide with a full memorized strain of up to ~14%. Moreover, a reversible recoverable strain of ~12±2% can be achieved by a single-step thermal activation or electric field with a large volumetric work density of up to 600±90 $J/cm^3$. These results show that a large SME at nanoscale can be obtained in an oxide material. Provided that the technical problems in devices such as hysteresis, fatigue, microcracking etc. can be avoided, this promising oxide material at its nanoscale may be potentially useful in nano-medical, energy storage, or nano-damping applications [36-37].

**Results**

   **Preparation of the unconstrained BFO structures.** BFO films have been grown on LaAlO$_3$ (LAO) substrates with a thin (La,Sr)CoO$_3$ (LSCO) bottom electrode (~5 nm), showing a nanoscale mixture of R/T phases. Direct observation of the thickness-dependent topography indicates that the R/T mixture has a highly strain-sensitive nature, as shown in the Supplementary Fig. S1. Interestingly, external stimuli such as electrical/thermal activation can induce an interface movement (see Supplementary Figs S2 and S3), resembling a martensitic transformation in traditional SMAs. This first glimpse gives us a strong impetus to explore the SME in this oxide. One good way to study its complete phase transformation-related shape deformation and memory effect is the manipulation of its stress distribution on a clamped film with a phase mixture, which can be tailored by reducing the lateral size of the film using focused ion beam (FIB) milling. Fig. 1(a) shows the typical image of the ion-milled BFO structures with a thickness of ~150 nm in order to reduce the clamping from the substrate. A robust piezoresponse force microscopy (PFM) based switching (see Supplementary Fig. S4 (a) and (b)) on such a FIB-milled sample indicates that ferroelectricity is maintained in this BFO structure. Before we study the stress-dependent structural transformation, we use continuum mechanics approaches to simulate the constraint profile of the BFO films as a function of lateral size [38-39]. The effective stress distribution ($\sigma_{\text{effective}}$) in the horizontal position ($x$) of the film is mainly determined by the aspect ratio ($l, h$), elastic constants of the film and substrate ($E_s$ and $E_f$), Poisson ratio ($v$), and piezoelectric susceptibility ($d_{31}$), applied electric field ($E_{\text{app}}$), etc., as in the following description (1) and (2):

$$\sigma_{\text{effective}} = \frac{E_f}{1-v} \times (1 - e^{k(1-x)}) \times d_{31} E_{\text{app}} \qquad (1)$$

$$k = \sqrt{\frac{3E_s(1-v_f)}{2E_f(1+v_s)h_f h_s}} \qquad (2)$$

Parameters for R- and/or T-BFO can be found elsewhere [40-41]. The degree of clamping of BFO as a function of lateral size has been estimated as shown in Fig. 1 (b). The clamping from the substrate has been estimated to decrease for lateral dimensions around 10 μm and is mostly removed around 1 μm. The inset shows the relative stress distribution within the whole etched structure with different sizes. In order to probe the piezoelectric coefficient of such an unclamped ferroelectric thin film, a capacitor structure was fabricated with Pt top electrodes. Fig. 1 (c) shows the $d_{33}$ as a function of AFM tip DC bias of pure R-phase BFO, constraint R/T BFO, and unconstraint R/T BFO capacitors with a lateral size of 1 μm. We can observe a significant enhancement of the $d_{33}$ of ~275±40 pm/V in this stress-released capacitor, which is about twice the value of full-clamped capacitors (~110±20 pm/V). Fig. 1(d) gives the phase loop during the measurement of the piezoresponse, indicative of a reversible ferroelectric switching process. The hysteresis phase loop and the complete piezo-amplitude curves (see Supplementary Fig. S4 (c)) indicate that the high $d_{33}$ can be reversibly obtained by sweeping the DC electric field. This large piezoresponse further confirms that the mechanical constraint from the substrate has been reduced with the decrease of lateral size of the capacitor [39].

**Size-dependent structural variation.** In order to study the structural transformation with the reduction of substrate clamping, AFM and micro-Raman spectra as a function of the lateral size of the ~150 nm thick BFO film (among the largest thickness before the structure relaxes) are shown in Fig. 2. Fig. 2(a)-(d) show the surface topography of the BFO film in the fully clamped state (as-grown state) and lateral sizes of 10 μm, 5 μm and 1 μm respectively, as illustrated in the corresponding schematic insets. As seen in these AFM images, the stripe-like structure in Fig. 2 (a) is a typical characteristic of the R/T mixture which is extensively studied elsewhere [28]. Based on this knowledge, we imaged the topography of the structure with small sizes etched by FIB. The stripe-like feature starts to change when the lateral size is smaller than 10 μm, and gradually fades away on the 5 μm and 1 μm structures as seen in Fig. 2 (c) and (d). The dissolving stripe pattern typical for phase mixture gives us a first glimpse of the structural variation and the SME associated with the stress alteration of the BFO. However, the AFM results cannot help make a conclusive argument for the detailed phase transformation, namely, has the phase mixture transformed to pure R-phase or T-phase with reduction of the clamping?

We performed micro-Raman studies, as this technique has been shown to be sensitive to the different structural symmetries related to BFO [42-44]. Measurements were performed on fully constrained films and FIB-milled structures with various sizes from 20 μm to 5 μm. Furthermore, we took advantage of the micro-Raman technique to characterize the structure of the FIB-milled R/T BFO film before and after electric field switching using a scanning probe setup. As shown in Fig. 2 (e), the micro-Raman spectra for different sizes exhibit peaks at 223 $cm^{-1}$, 269 $cm^{-1}$, and 364 $cm^{-1}$ indicating the T-phase [44]. The intensity of these peaks is found to decrease for smaller size showing a decrease in the fraction of the T-phase. Moreover, after a poling (4×4 $μm^2$ box in the 5 μm structure) by +20 V (~1.2 MV/cm), the same Raman peaks at 223 $cm^{-1}$, 269 $cm^{-1}$, and 364 $cm^{-1}$ are found to increase in intensity accompanying the big electric-field-induced SME. This indicates the presence of a larger fraction of the T-phase as compared to the unswitched sample. In constrained films, the peak positions for the R- and T-phase deviated far from the respective pure phases due to the substrate clamping effect. For this reason we can only observe a relatively low strain (less than 5%) with the external stimuli such as electric field or temperature in a constrained film as shown in Supplementary Figs S2 and S3.

**Clamping-dependent phase transformation.** The most straightforward way to observe the phase transformation of thin films in association with lateral clamping is through TEM. Therefore High Angle Annular Dark Field Scanning Transmission Electron Microscopy (HAADF-STEM) together with selected area electron diffraction (SAED) was performed on the ~150 nm thick BFO with the R/T mixture. A thin lamella of the cross-sectional sample was prepared by FIB milling. As seen in the Supplementary Fig. S5, an area showing an array of stripe-like R/T mixture (region in the brown box) is milled out together with its substrate by FIB. In order to study its cross-sectional structure with a size gradient (to gradually reduce the degree of clamping from the substrate), we further milled this sliced electron-transparent sample into a wedge shape where the lateral thickness varies from ~300 nm down to ~50 nm as seen in the inset (a top view) of Fig. 3 (a).

The HAADF-STEM image of the as-prepared lamella is shown in Fig. 3. The typical stripe pattern of the R/T mixture is clear in the cross-sectional view of the 300 nm thick area as shown in the right side in Fig. 3 (a): the dark area indicates the T-BFO while the bright area indicates the R-BFO. From the cross-sectional view of the wedge-shaped area (indicated by yellow box in the inset of Fig. 3 (a)), we can still observe that the stripe-like feature remains near the 300 nm thick area and gradually disappears with the decrease of the thickness down to 50 nm (in the left side in Fig. 3 (a)) in the TEM sample, which indicates a structural transformation from R/T mixture to a pure phase with less substrate clamping. We can also observe that the phase relaxation starts from the top of the sample surface and down to the bottom of the film with the reduction of clamping, which is consistent with previous size-dependent AFM studies. The detailed atomic structure in the strain-relaxed region (50 nm thick) is shown in Fig. 3 (b), where a high-resolution TEM image reveals an out-of-plane lattice constant of 4.06 Å (calibrated to the LAO substrate), which clearly denotes the R-phase. This phase structure has been further confirmed by its corresponding SAED pattern in the inset. The phase boundary with R/T mixture is also confirmed by an atomic-resolution HAADF-STEM image shown in Fig. 3 (c).

The TEM results have shown that a SME from R/T mixture to pure R-phase in BFO can be tailored by its stress distribution. If the surface displacement associated with this martensitic-like structural transformation is carefully measured from the TEM image, we can directly observe a remnant strain of ~9% (shape memory from R/T mixture to pure R-phase). Along with the electric-field-/temperature-induced strain of ~5% (shape memory from R/T mixture to pure T-phase, see supplementary Fig. S2 and S3), we are able to obtain a full memorized strain of up to ~14%.

**In-situ TEM.** In order to accomplish a full cycle of the structural transformation that is analogous to the one found in traditional SMAs, the key issue would be whether this structural deformation or shape memory can be recovered. We use in-situ TEM with a temperature-controlled holder to demonstrate the repeatable cycling of this large SME. Fig. 4 (a) and (c) show the TEM images and the diffraction pattern of relaxed pure R-BFO. When the temperature reaches 573 K, the T-phase starts to emerge, with a stripe-like phase coexistence as shown in Supplementary Fig. S6. When the oxide is further heated up to 673 K, the pure T-phase is completely recovered with a significant shape deformation (Fig. 4 (b)) and an out-of-plane lattice constant of ~4.6 Å confirmed from the diffraction pattern (Fig. 4 (d)). Relative to the relaxed R-BFO with an out-of-plane lattice constant of ~4.06 Å (obtained from diffraction pattern in Fig. 4(c)), the average recoverable strain can reach ~12±2% (as seen in Fig. 4 (b)) through temperature cycle. The reversibility of the structural transformation with a significant strain in this oxide is summarized in Fig. 4 (e), where the corresponding images can be seen in Supplementary Fig. S7, showing a single-step control of this large reversible SME. An animation of a nano-actuator with a large switchable deformation through temperature control is demonstrated as seen in supplementary movie 1.

To fulfill a room-temperature device with a high response frequency, we can also observe this large reversible SME (>10%) in the unconstrained structure driven by a moderate electric field of 1-2 MV/cm, as shown in Supplementary Fig. S8. This effect is absent in an unconstrained PZT island structure (see Supplementary Fig. S9). Based on our full understanding of the electric field/stress and temperature of structural transformation, we are able to draw a quasi-quantitative deformation/recovery cycle attributed to electric field/temperature and stress, as shown in Fig. 5 (a). This resembles the "De-twinning", "Heating", and "Cooling" processes attributed to thermal and stress stimuli in traditional SMAs [1-2]. However, in metal-based alloys at the nanoscale, the surface and interfacial free energy play critical roles in the thermodynamic phase equilibrium and kinetic aspects of heterogeneous nucleation and growth, which means that a complex competition of strain/interface energy are a prerequisite to fulfill such a phase transformation [11-16]. In our oxide nano-system, the highly strained BFO film provides a good model system to realize the nanoscale control between elastic transformation, electric, and interfacial energies, resulting in a SME with a large strain. Besides, the volumetric work density ($E_{elastic} \cdot \varepsilon_{max}^2/2$) are dependent on Young's modulus ($E_{elastic}$) and the maximum strain ($\varepsilon_{max}$). The Young's modulus (70±10 GPa) of the film was assumed to be constant during the phase transformation (the mechanical test of the material can be seen in Supplementary Fig. S10), which provides an upper bound of 600±90 J/cm$^3$ for the average volumetric work density as it is probable that the material softens during the transformation. This is almost two orders of magnitude larger than the one of traditional SMAs, electrostrictive polymers [45] etc.

## Discussion

The physical origin of the martensitic-like shape memory behavior in this nanoscale oxide material and its large remnant strain has been further investigated by first principles calculations. We have studied the interplay between structural transformation, polarization and tetragonality (c/a ratio) in a strained BFO, and their coupling to external electric field and mechanical stress.

Two stable phases, namely R-phase and T-phase, are found in BFO under compressive strain. The strain-dependent properties for both R- and T-phase of BFO are shown in Supplementary Fig. S11. A martensitic-like phase transformation can occur when compressive strain is about -5%, accompanied by a significant enhancement of c/a ratio (1.08 for R and 1.26 for T) and out-of-plane polarization (P [001] = 0.82 for R-BFO and 1.41 C/m$^2$ for T-BFO). These are consistent with previous results [46-47]. We further investigated how the structural deformation and transformation occurs under our experimental conditions. Fig. 5 (b) shows the evolution of the energy as a function of c/a ratio under a strain of -5.8%. The two energy minima correspond to the stable R- and T-phase respectively. The double-well-like energy curve suggests that both T- and R-phase can coexist in BFO, which is in good agreement with the R/T mixture observed experimentally. By reducing the degree of clamping from the substrate, BFO films under smaller compressive strain will transform to the stable R-phase, instead of the metastable T-phase and the transition from R- to T-phase will be unfavorable (Fig. 5 (c)). Therefore, the transformation from the R/T mixture to the pure R-phase can be realized with the reduction of substrate clamping. By applying the electric field along the [001] direction (Fig. 5 (d)), the stability of the T-phase can be largely enhanced [48]. Again, the electric-field-induced SME can occur as a full cycle, accompanied by the structural transformation. We also obtain the exact relation between polarization and axial c/a ratio. Similar to tetragonal $PbTiO_3$ solid solutions [49], we found that $P_z^2$ is proportional to the tetragonality for both R- and T-phase around the transition region (inset of Supplementary Fig. S11). Such a strong coupling between polarization and c/a ratio also indicates that the abrupt change of out-of-plane polarization is always accompanied with strong structural deformations, which further suggests that such a large SME in this oxide is electrically controllable.

In principle, there should be no SME exhibited between R and T phases in oxide compounds such as $(1-x)[Pb(Mg_{1/3}Nb_{2/3})O_3]-x[PbTiO_3]$ and $Pb(Zr,Ti)O_3$, due to their low symmetry nature with more variants at the same energy level. However, mechanical stress can favor a certain structural variant in elastic systems, which helps stabilize the pure R- and T-phase, as well as the R/T mixture in BFO. Therefore, a large shape deformation can be memorized due to the structural transformation from the interplay between stress, electric fields and even temperature [50]. Another important information from the calculated result is that T-phase has a significantly higher energy than R-phase under a strain-relaxed state. This indicates that the transition from R to T is difficult under such a condition, and this may explain why BFO is a very "hard" material under zero stress, and it becomes "softer" at the presence of suitable clamping stress. Besides, if a tensile stress can be applied normal to the film surface, the energy profile of this system suggests that the full SME with a strain of up to ~14% can be also achieved by a single-step external mechanical stimulus.

To conclude, we have demonstrated a large recoverable memorized strain of up to 14% in BFO at the nanoscale, driven by the interplay between electric field, temperature and stress. Analogous to the effect in conventional alloys, such a large SME is for the first time observed in an oxide material. The martensitic-like phase transformation has been investigated by a combination of AFM, micro-Raman, TEM, and first principles calculations. A full cycling of the phase evolution can be also achieved by a single-step thermal activation with a reversible strain of ~12±2%. Electric field can also be used to fulfill this reversible shape memory at room temperature with a volumetric work density of up to 600±90 J/cm$^3$. This large SME coupled with the full insight of its nanoscale control, may have great potential for future applications in nano-architectures.

## Methods

**Detailed growth conditions.** Epitaxial BFO films with the mixture of T- and R-phase were grown on (001) LAO substrates by pulsed laser deposition (PLD) with (La,Sr)MnO$_3$ (LSMO) or (La,Sr)CoO$_3$ (LSCO) as bottom electrodes (~5 nm). The growth temperature was kept at 700 ℃ and the working oxygen pressure was 100 mTorr. For all the films, a growth rate of 2.5 nm/min and a cooling rate of 5 ℃/min under an oxygen atmosphere were used. A laser energy density of 1.2 J/cm$^2$ and repetition rate of 5 Hz were used during the deposition. The film thickness can be controlled as about 40-160 nm.

**Sample milling parameters.** Island structures for AFM characterizations were fabricated by 30 keV Ga+ ions in an FEI 235 dual beam focused ion beam (FIB). The ion beam current we applied for milling all of the samples was 50 pA. The milling time used was 2.5 min for the islands with diameters of 3 μm and 1 μm. 3.5 min was applied for milling the islands with diameters of 10 μm and 5 μm. An important issue for the FIB sample preparation is to protect the sample from damage caused by the incident 30 kV Ga+ ion beam. In our sample fabrication process, a 10 pA ion beam current (with 10 second milling time) was used at the final milling step to minimize the ion beam damage to the sample. The whole process has been done without any imaging.

**AFM/ PFM measurements.** Scanning probe-based measurement has been carried out on a Digital Instruments Nanoscope-IV Multimode AFM equipped with a conductive AFM application module under ambient conditions. Surface topography on the clamping-removed structures has been imaged by a contact-mode AFM using a sharp Si tip. Local piezoelectric properties were studied using a commercially available TiPt-coated Si tips (MikroMasch). We calibrate our PFM measurement by using a standard quartz crystal (its piezoelectric constant of 2.3 pm/V). For each measurement of piezoelectric susceptibility ($d_{33}$ curve), we use the scanning probe system to measure the quartz crystal once and then use exactly the same condition to measure our film and island capacitors for calibration. These are local measurements on small capacitors which are benefitted by our nanoscale tip. During the measurement, typical scan rates were 0.2 Hz. The ferroelectric domain structures have been captured at a 6.39 kHz with a tip-biased voltage of 0.5 $V_{pp}$. The poling voltage on the electrically biased tip was controlled up to 20 V (~1.2 MV/cm). Using the setup described above, high-resolution PFM measurements were completed on a wide array of samples before and after switching.

**Detailed measurement for Micro-Raman spectra.** Micro-Raman spectroscopy, which is a powerful tool to investigate the vibrational, rotational, and other low-frequency modes in a system, which can be used to determine the lattice structures corresponding to different Raman shifts, especially for structures with small sizes. Unpolarized local Raman spectra on FIB prepared structures were acquired using a micro-Raman setup at 476 nm wavelength with a spot size of 1-2 microns at room temperature. Integration times were of the order of 30 minutes for each spectrum. The light spot was carefully monitored to exclude the drift off the island during data acquisition.

**FIB milling of TEM samples.** The TEM samples were also prepared in a FIB process by a FEI Helios NanoLab DualBeam system. A thick lamella of ~1 μm in thickness and ~10 μm in lateral size is milled using 30 keV $Ga^+$ ions. For preparing a wedge-shaped thin lamella, further thinning was carried out using 16 keV and 8 keV $Ga^+$ ions using a cleaning-cross-sectional pattern with an inclined angle at each side. A final cleaning of the sample surface was performed using 5 keV and 2 keV $Ga^+$ ions with small beam current. The as-prepared thin lamella has a wedge shape with thinnest area of ~50 nm and thick area of ~300 nm.

**TEM imaging conditions.** High Resolution High Angle Annular Dark Field Scanning Transmission Electron Microscopy (HR-HAADFS-STEM) was carried out on FEI Titan Cube 50-80 microscope fitted with an aberration-corrector for the imaging lens and another for the probe forming lens as well as a monochromator, operated at 300 kV. The STEM convergence semi-angle used was ~21.4 mrad, providing a probe size of ~1.0 Å at 300 kV. Diffraction patterns were acquired at FEI Tecnai microscope operated at 200 kV. The In-situ TEM coupled with a temperature control has been done by a dedicated holder.

**Mechanical test on mixed-phase BFO.** The mechanical test on the mixed-phase BFO has been carried out in a TEM equipment coupled with a force probe with a contact area of ~ 5000 $nm^2$. The maximum strain value from a mixed-phase to pure R phase is ~10%. The plotted Young's modulus is $70\pm10$ GPa during the phase transformation ($\sigma/\varepsilon$), which is consistent with previous experimental and theoretical resuts [51-53]. The corresponding diffraction patterns have been recorded during the indentation in order to keep track of the phase transformation at each stage.

**First-principles calculations.** In our first principles calculations with Quantum Espresso (QE) code, we use norm-conserving [54] optimized designed nonlocal pseudopotentials [55]. Partial core corrections (PCC) [56] are included in the Fe pseudopotential. Our calculations are performed using the local spin-density approximation (LSDA) plus Hubbard U approach as implemented in QE [57], with the effective U value of 3.8 eV applied on Fe [58]. The Fe ions are in the G-type ordered antiferromagnetic state (G-AFM), and the spin-orbital coupling is not included. We sample the BZ using an 8×8×8 Monkhorst-Pack $k$-point mesh. A plane-wave energy cut-off 60 Ry was used for calculation. The electronic contribution to the polarization is calculated following Berry phase formalism [59,60]. A homogeneous finite electric field applied on BFO thin films is described through the modern theory of the polarization [61]. We use the "strained bulk" simulation [62] to represent the strained BFO thin films. All strain values are given relative to the bulk LDA+U pseudo-cubic lattice parameter (3.965 Å). A 20-atom super-cell, with lattice vector $l_1$=(a, a, 0), $l_2$=(a, $\underline{a}$, 0) and $l_3$=(0, 0, 2c), is used for simulation. Epitaxial strain is introduced by the mismatch of pseudocubic lattice constants between the substrate and BFO film. To simulate the epitaxial strained BFO, the (001)-oriented cubic substrate with an in-plane lattice constant (a) is fixed, while out-of-plane lattice parameter (c) and the monoclinic angle (the angle between lattice $l_1$ and $l_3$) are free to relax. The atomic positions in the BFO structure are optimized until the Hellmann-Feynman forces on the atoms are less than 0.1 meVÅ$^{-1}$.

# References:

[1] X. Ren and K. Otsuka, Origin of rubber-like behaviour in metal alloys. *Nature* **389,** 579-582 (1997).

[2] R. W. Cahn, Nature Matallic ruber bounces back. *Nature* **374,** 120-121 (1995).

[3] K. Bhattacharya, S. Conti, G. Zanzotto, and J. Zimmer, Crystal symmetry and the reversibility of martensitic transformations. *Nature* **428,** 55-59 (2004).

[4] N. B. Morgan, Medical shape memory alloy applications. *Mater. Sci. Eng.* **A378,** 16-23 (2004).

[5] J. Van Humbeeck, Non-medical applications of shape memory alloys. *Mater. Sci. Eng.* **A 273,** 134-148 (1999).

[6] S. Saadat, J. Salichs, M. Noori, Z. Hou, H. Davoodi, I. Bar-on, Y. Suzuki, and A. Masuda, An overview of vibration and seismic applications of NiTi shape memory alloy. *Smart Mater. Struct.* **11,** 218-229 (2002).

[7] K. Otsuka and X. Ren, Physical metallurgy of Ti-Ni-based shape memory alloys. *Progr. Mater. Sci.* **50,** 511-678 (2005).

[8] R. Kainuma, Y. Imano, W. Ito, Y. Sutou, H. Morito, S. Okamoto, O. Kitakami, K. Oikawa, A. Fujita, T. Kanomata, and K. Ishida, Magnetic-field-induced shape recovery by reverse phase transformation. *Nature* **439,** 957-960 (2006).

[9] M. Chmielus, X. X. Zhang, C. Witherspoon, D. C. Dunand, and P. Müllner, Giant magnetic-field-induced strains in polycrystalline Ni–Mn–Ga foams. *Nat. Mater.* **8,** 863-866 (2009).

[10] K. Ullakko, J. K. Huang, C. Kantner, R. C. O'Handley, and V. V. Kokorin, Large magnetic−field−induced strains in $Ni_2MnGa$ single crystals. *Appl. Phys. Lett.* **69,** 1966-1968 (1996).

[11] T. Waitz, K. Tsuchiya, T. Antretter, and F. D. Fischer, Phase transformations of nanocrystalline martensitic materials. *MRS Bulletin* **34,** 814-821 (2009).

[12] T. Waitz, V. Kazykhanov, and H. P. Karnthaler, Martensitic phase transformations in nanocrystalline NiTi studied by TEM. *Acta Mater.* **52,** 137-147 (2004).

[13] K. Seki, H. Kura, T. Sato, and T. Taniyama, Size dependence of martensite transformation temperature in ferromagnetic shape memory alloy FePd. *J. Appl. Phys.* **103,** 063910 (2008).

[14] H. S. Yang and H. K. D. H. Bhadeshia, Austenite grain size and the martensite-start temperature. *Scripta Mater.* **60,** 493-495 (2009).

[15] M. Bouville and R. Ahluwalia, Microstructure and mechanical properties of constrained shape-memory alloy nanograins and nanowires. *Acta Mater.* **56,** 3558-3567 (2009).

[16] F. D. Fischer, T. Waitz, D. Vollath, and N. K. Simha, On the role of surface energy and surface stress in phase-transforming nanoparticles. *Progr. Mater. Sci.* **53,** 481-527 (2008).

[17] V. Hardy, S. Majumdar, S. J. Crowe, M. R. Lees, D. McK. Paul, L. Herve, A. Maignan, S. Hebert, C. Martin, C. Yaicle, M. Hervieu, and B. Raveau, Field-induced magnetization steps in intermetallic compounds and manganese oxides: The martensitic scenario. *Phys. Rev. B* **69,** 020407(R) (2004).

[18] V. Podzorov, B. G. Kim, V. Kiryukhin, M. E. Gershenson, and S-W. Cheong, Martensitic accommodation strain and the metal-insulator transition in manganites. *Phys. Rev. B* **64,** 140406(R) (2001).

[19] N. N. Loshkareva, D. I. Gorbunov, A. V. Andreev, N. V. Mushnikov, Y. Skourski, and F. Wolff-Fabris, Metamagnetic transition of martensitic type in electron-doped manganites $Ca_{1-x}Ce_xMnO_3$ (x = 0.10, 0.12). *J. Alloy Comp.* **553,** 199-203 (2013).

[20] N. Navruz, Crystallographic features of the martensitic transformation in $PbTiO_3$ compound. *AIP Conf. Proc.* **1203,** 862-865 (2010).

[21] H. Fu and R. E. Cohen, Polarization rotation mechanism for ultrahigh electromechanical response in single-crystal piezoelectrics. *Nature* **403,** 281-283 (2000).

[22] S. R. Anton and H. A. Sodano, A review of power harvesting using piezoelectric materials. Smart Mater. Struct. 16, R1-R21 (2007).

[23] S. H. Baek, J. Park, D. M. Kim, V. A. Aksyuk, R. R. Das, S. D. Bu, D. A. Felker, J. Lettieri, V. Vaithyanathan, S. S. N. Bharadwaja, N. Bassiri-Gharb, Y. B. Chen, H. P.

Sun, C. M. Folkman, H. W. Jang, D. J. Kreft, S. K. Streiffer, R. Ramesh, X. Q. Pan, S. Trolier-McKinstry, D. G. Schlom, M. S. Rzchowski, R. H. Blick, and C. B. Eom, Giant piezoelectricity on Si for hyperactive MEMS. *Science* **334,** 985-961 (2011).

[24] S. –E. Park and T. R. Shrout, Ultrahigh strain and piezoelectric behavior in relaxor based ferroelectric single crystals. *J. Appl. Phys.* **82,** 1804-1811, (1997).

[25] W. Liu and X. Ren, Large piezoelectric effect in Pb-free ceramics. *Phys. Rev. Lett.* **103,** 257602 (2009).

[26] Z. Kutnjak, J. Petzelt, and R. Blinc, The giant electromechanical response in ferroelectric relaxors as a critical phenomenon. *Nature* **441,** 956-959 (2006).

[27] X. Ren, Large electric-field-induced strain in ferroelectric crystals by reversible domain switching. *Nat. Mater.* **3,** 91-94 (2004).

[28] R. J. Zeches, M. D. Rossell, J. X. Zhang, A. J. Hatt, Q. He, C.-H. Yang, A. Kumar, C. H. Wang, A. Melville, C. Adamo, G. Sheng, Y.-H. Chu, J. F. Ihlefeld, R. Erni, C. Ederer, V. Gopalan, L. Q. Chen, D. G. Schlom, N. A. Spaldin, L. W. Martin, and R. Ramesh, A strain-driven morphotropic phase boundary in $BiFeO_3$. *Science* **326,** 977-980 (2009).

[29] Z. Chen, Z. Luo, C. Huang, Y. Qi, P. Yang, L. You, C. Hu, T. Wu, J. Wang, C. Gao, T. Sritharan, and L. Chen, Low-symmetry monoclinic phases and polarization rotation path mediated by epitaxial strain in multiferroic $BiFeO_3$ thin films. *Adv. Funct. Mater.* **21,** 133-138 (2011).

[30] J. C. Yang, Q. He, S. J. Suresha, C. Y. Kuo, C. Y. Peng, R. C. Haislmaier, M. A. Motyka, G. Sheng, C. Adamo, H. J. Lin, Z. Hu, L. Chang, L. H. Tjeng, E. Arenholz, N. J. Podraza, M. Bernhagen, R. Uecker, D. G. Schlom, V. Gopalan, L. Q. Chen, C. T. Chen, R. Ramesh, and Y. H. Chu, Orthorhombic $BiFeO_3$. *Phys. Rev. Lett.* **109,** 247606 (2012).

[31] R. K. Vasudevan, Y. Liu, J. Li, W. I. Liang, A. Kumar, S. Jesse, Y. C. Chen, Y. H. Chu, V. Nagarajan, and S. V. Kalinin, Nanoscale Control of Phase Variants in Strain-Engineered $BiFeO_3$. *Nano Lett.* **11,** 3346-3354 (2011).

[32] Q. He, Y. H. Chu, J. T. Heron, S. Y. Yang, W. I. Liang, C. Y. Kuo, H. J. Lin, P. Yu, C. W. Liang, R. J. Zeches, W. C. Kuo, J. Y. Juang, C. T. Chen, E. Arenholz, A. Scholl,

and R. Ramesh, Electrically controllable spontaneous magnetism in nanoscale mixed phase multiferroics. *Nat. Commun.* **2,** 225-229 (2011).

[33] A. R. Damodaran, C.-W. Liang, Q. He, C.-Y. Peng, L. Chang, Y.-H. Chu, and L. W. Martin, Nanoscale structure and mechanism for enhanced electromechanical response of highly strained $BiFeO_3$ thin films. *Adv. Mater.* **23,** 3170-3175 (2011).

[34] J. X. Zhang, B. Xiang, Q. He, J. Seidel, R. J. Zeches, P. Yu, S. Y. Yang, C. H. Wang, Y-H. Chu, L. W. Martin, A.M. Minor, and R. Ramesh, Large field induced strains in a lead free piezoelectric material. *Nat. Nanotechnol.* **6,** 98-102 (2011).

[35] S. Lisenkov, D. Rahmedov, and L. Bellaiche, Electric-Field-Induced Paths in ... from Atomistic Simulations. *Phys. Rev. Lett.* **103,** 047204 (2009).

[36] Jose San Juan, M. L. Nó, and C. A. Schuh, Thermomechanical behavior at the nanoscale and size effects in shape memory alloys. *J. Mater. Res.* **26,** 2461-2469 (2011).

[37] Jose San Juan, M. L. Nó, and C. A. Schuh, Nanoscale shape-memory alloys for ultrahigh mechanical damping. *Nat. Nanotechnol.* **4,** 415-419 (2009).

[38] E. Suhir, An approximate analysis of stresses in multilayered elastic thin films. *J. Appl. Mech.* **55,** 143-148 (1988).

[39] V. Nagarajan, A. Roytburd, A. Stanishevsky, S. Prasertchoung, T. Zhao, L. Chen, J. Melngailis, O. Auciello, and R. Ramesh, Dynamics of ferroelastic domains in ferroelectric thin films. *Nat. Mater.* **2,** 43-47 (2003).

[40] S. A. T. Redfern, Can Wang, J. W. Hong, G. Catalan, and J. F. Scott, Elastic and electrical anomalies at low-temperature phase transitions in $BiFeO_3$. *J. Phys.: Condens. Matter* **20,** 452205 (2008).

[41] C. Ederer and N. A. Spaldin, Effect of epitaxial strain on the spontaneous polarization of thin film ferroelectrics. *Phys. Rev. Lett.* **95,** 257601 (2005).

[42] M. O. Ramirez, A. Kumar, S. A. Denev, Y. H. Chu, J. Seidel, L. W. Martin, S.-Y. Yang, R. C. Rai, X. S. Xue, J. F. Ihlefeld, N. J. Podraza, E. Saiz, S. Lee, J. Klug, S-. W. Cheong, M. J. Bedzyk, O. Auciello, D. G. Schlom, J. Orenstein, R. Ramesh, J. L. Musfeldt, A. P. Litvinchuk, and V. Gopalan, Spin-charge-lattice coupling through resonant multimagnon excitations in multiferroic $BiFeO_3$. *Appl. Phys. Lett.* **94,**

161905 (2009).

[43] D. Kothari, V. Raghavendra Reddy, V. G. Sathe, A. Gupta, A. Banerjee, and A. M. Awasthi, Raman scattering study of polycrystalline magnetoelectric BiFeO$_3$. *J. Magn. Magn. Mater.* **320,** 548-552 (2008).

[44] M. Iliev, M. V. Abrashev, D. Mazumdar, V. Shelke, and A. Gupta, Polarized Raman spectroscopy of nearly tetragonal BiFeO$_3$ thin films. *Phys. Rev. B* **82,** 014107 (2010).

[45] K. Liu, C. Cheng, Z. Cheng, K. Wang, R. Ramesh, and J. Wu, Giant-amplitude, high-work density microactuators with phase transition activated nanolayer bimorphs. *Nano Lett.* **12,** 6302-6308 (2012).

[46] J. C. Wojdel and J. Iniguez, *Ab Initio* indications for giant magnetoelectric effects driven by structural softness. *Phys. Rev. Lett.* **105,** 037208 (2010).

[47] A. J. Hatt, N.A. Spaldin, and C. Ederer, Strain-induced isosymmetric phase transition in BiFeO$_3$. *Phys. Rev. B* **81,** 054109 (2010).

[48] S. Lisenkov, D. Rahmedov, and L. Bellaiche, Electric-field-induced paths in multiferroic BiFeO$_3$ from atomistic simulations. *Phys. Rev. Lett.* **103,** 047204 (2009).

[49] T. Qi, I. Grinberg, and A.M. Rappe, Correlations between tetragonality, polarization, and ionic displacement in PbTiO$_3$-derived ferroelectric perovskite solid solutions. *Phys. Rev. B* **82,** 134113 (2010).

[50] D. Mutter and P. Nielaba, Simulation of the thermally induced austenitic phase transition in NiTi nanoparticles simulation of phase transitions in NiTi nanoparticles. *Eur. Phys. J. B* **84,** 109-113 (2011).

[51] S. L. Shang, G. Sheng, Y. Wang, L. Q. Chen, and Z. K. Liu, Elastic properties of Cubic and Rhombohedral BiFeO$_3$ from first-principles calculations. Phys. Rev. B **80,** 052102 (2009).

[52] A. G. Gavriliuk, V. V. Struzhkin, I. S. Lyubutin, S. G. Ovchinnikov, M. Y. Hu, and P. Chow, Another mechanism for the insulator-metal transition observed in Mott insulators. Phys. Rev. B **77,** 155112 (2008).

[53] S. A. T. Redfern, C.Wang, J. W.Hong, G. Catalan, and J. F. Scott, Elastic and electrical anomalies at low-temperature phase transitions in BiFeO$_3$. J. Phys.:

Condens. Matter **20,** 452205 (2008).

[54] A. M. Rappe, K. M. Rabe, E. Kaxiras, and J. D. Joannopoulos, Optimized pseudopotentials. Phys. Rev. B **41,** 1227-1230 (1990).

[55] N. J. Ramer and A. M. Rappe, Designed nonlocal pseudopotentials for enhanced transferability. Phys. Rev. B **59,** 12471-12478 (1999).

[56] S. G. Louie, S. Froyen, and M. L. Cohen, Nonlinear ionic pseudopotentials in spin-density-functional calculations. Phys. Rev. B **26,** 1738-1742 (1982).

[57] M. Cococcioni and S. de Gironcoli, Linear response approach to the calculation of the effective interaction parameters in the LDA+U method. Phys. Rev. B **71,** 035105 (2005).

[58] I. A. Kornev, S. Lisenkov, R. Haumont, B. Dkhil, and L. Bellaiche, Finite-temperature properties of multiferroic $BiFeO_3$. Phys. Rev. Lett. **99,** 227602 (2007).

[59] R. D. King-Smith and D. Vanderbilt, Theory of polarization of crystalline solids. Phys. Rev. B **47,** 1651-1654 (1993).

[60] D. Vanderbilt and R.D. King-Smith, Electric polarization as a bulk quantity and its relation to surface charge. Phys. Rev. B **48,** 4442 (1993).

[61] I. Souza, J.Iniguez, and D.Vanderbilt, First-principles approach to insulators in finite electric fields. Phys. Rev. Lett. **89,** 117602 (2002).

[62] O. Dieguez, K.M. Rabe, and D. Vanderbilt, First-principles study of epitaxial strain in perovskites. Phys. Rev. B **72,** 144101 (2005).

# Figure Captions

**Figure 1 | A way to remove the substrate clamping.** (a) a 3-D view of the typical FIB-milled structure in BFO, (b) an estimation of the degree of clamping from the substrate on the BFO film with R/T mixture as a function of the lateral size of the milled film, (c) quantitative measurements of the effective piezoelectric coefficient on the constraint and unconstraint BFO capacitors, (d) a ferroelectric phase switching loop from a unconstraint R/T BFO capacitor.

**Figure 2 | Structure of an unconstraint BFO film with R/T mixture.** AFM images of the BFO thin film with stripe-like phase boundaries before FIB milling (a), and the FIB-milled 10 μm size (b), 5 μm size (c) and 1 μm size (d), from which we can observe the stripe-like phase boundaries on the surface gradually vanish with the removal of clamping; (e) micro-Raman spectra on a fully clamped BFO film with the phase mixture and the corresponding unconstrained structures, while the variation of the relative intensity gives a first glimpse about the structural transformation from T-BFO to R-BFO with the removal of clamping. The scale bar is 2 μm in (a)-(c) and 200 nm in (d).

**Figure 3 | A direct evidence of the phase transformationby stress tailoring.** (a) Cross-sectional HAADF-STEM image of the phase evolution (the stripe-like areas labeled by yellow arrows indicates the T/R phase mixture) with a gradual release of the in-plane stress from right to left (the inset is the top view of this electron-transparent TEM sample), demonstrating a remnant strain of ~9% by achieving the structural transformation, (b) high-resolution image and the corresponding diffraction pattern from a fully stress-relaxed area (50 nm thick) of the sample showing a pure R-phase BFO, (c) high-resolution image from a relative clamped area (300 nm thick) of the sample near a R/T phase boundary. The scale bar is 100 nm in (a) and 2 nm in (b) and (c).

**Figure 4 | A structural recovery by heating with a large reversible strain.** (a) A low-magnification image of the relaxed pure R-BFO with protective coating layers, (b) A low-magnification image of the recovered pure T-BFO with an observable shape change of ~14% at 673 K, (c) the structural evidence of the relaxed pure R-BFO with a c-axis lattice constant of ~4.06 Å, (d) the structural evidence of the recovered pure T-phase BFO with a c-axis lattice constant of ~4.65 Å by an in-situ heating treatment, (e) the reversibility of the strain of ~12±2% by temperature cycling, where the error bars are from multiple measurements of the shape deformation on different locations of the sample (the scale bar is 100 nm).

**Figure 5 | An analogue of the full shape memory cycle**. (a) A complete cycle can be obtained by a combination of electrical/stress/thermal stimuli, accompanying a large recoverable strain of up to ~14% due to the structural transformation from a martensitic-like (M) R-phase to an austenite-like (A) T-phase. Afterwards, a one-step control with a recoverable strain of ~12±2% can be reversibly achieved by a thermal or electric field activation between both structures, (b) and (c) calculated energy of BFO as a function of c/a ratio at -5.8% and -4% strains, respectively, (d) calculated energy of BFO under a homogeneous electric field of 0.2 V/nm along [001] direction. In (b)-(d), the energy of the ground-state structure is set as energy zero, in good agreement with our experimental observations.

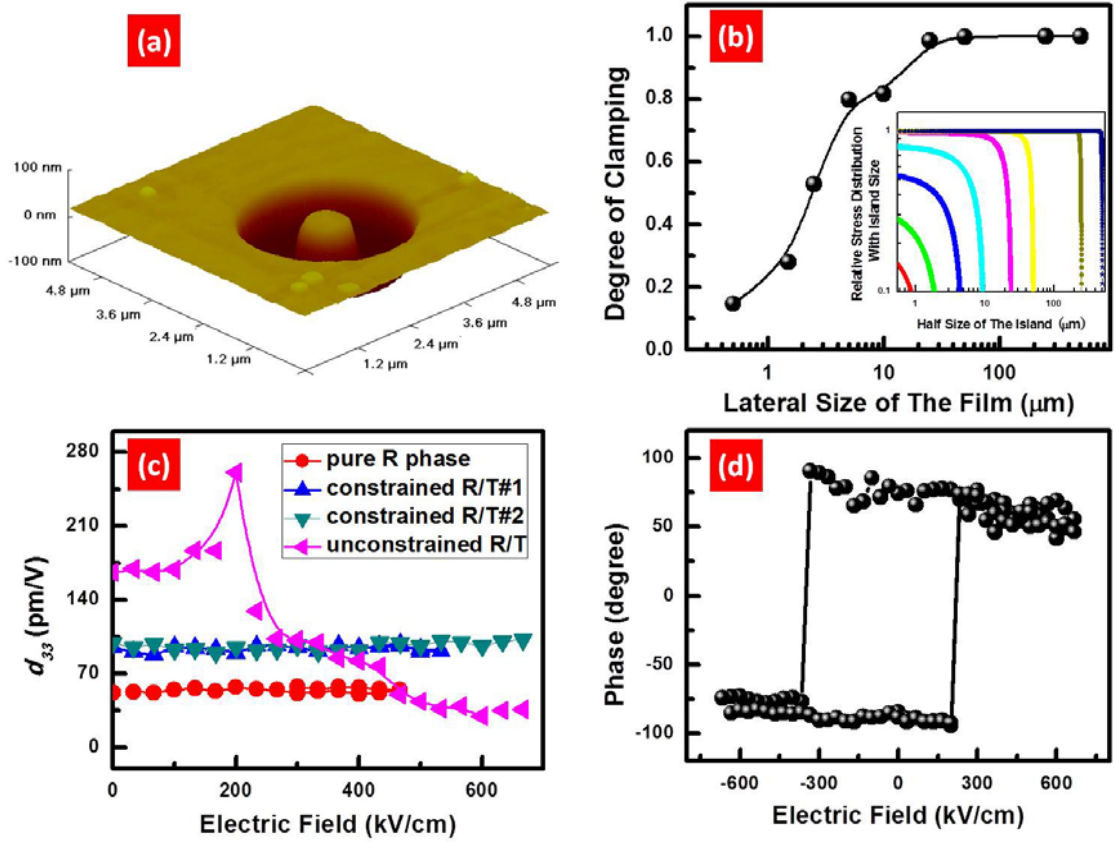

**Fig. 1. Zhang et al.**

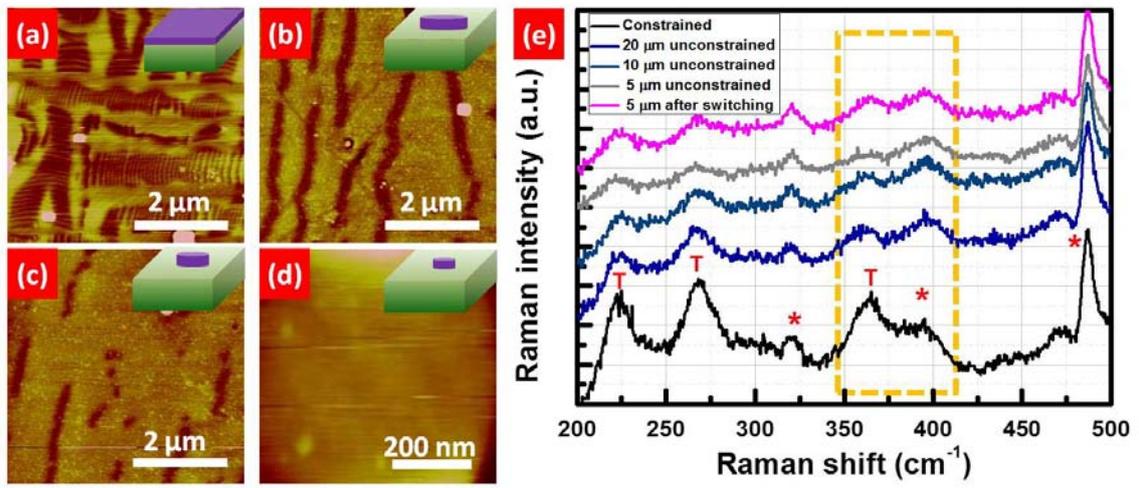

**Fig. 2.** Zhang et al.

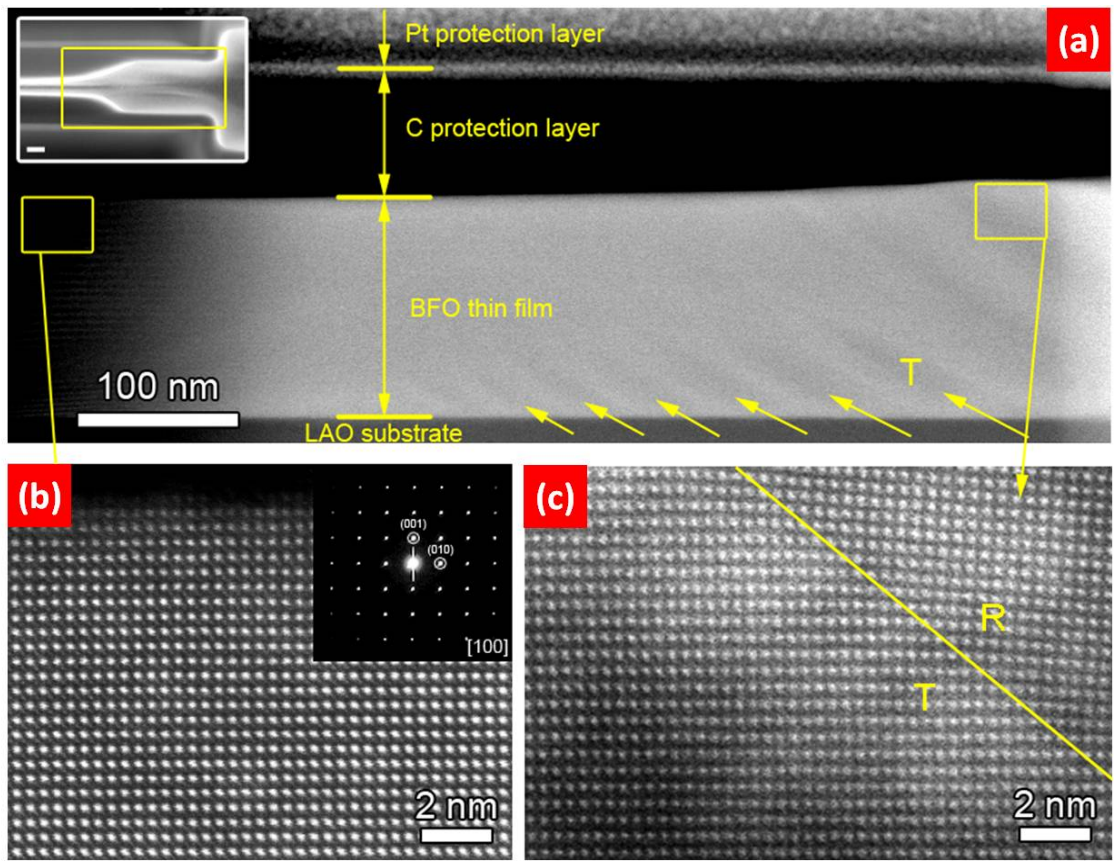

**Fig. 3.** Zhang et al.

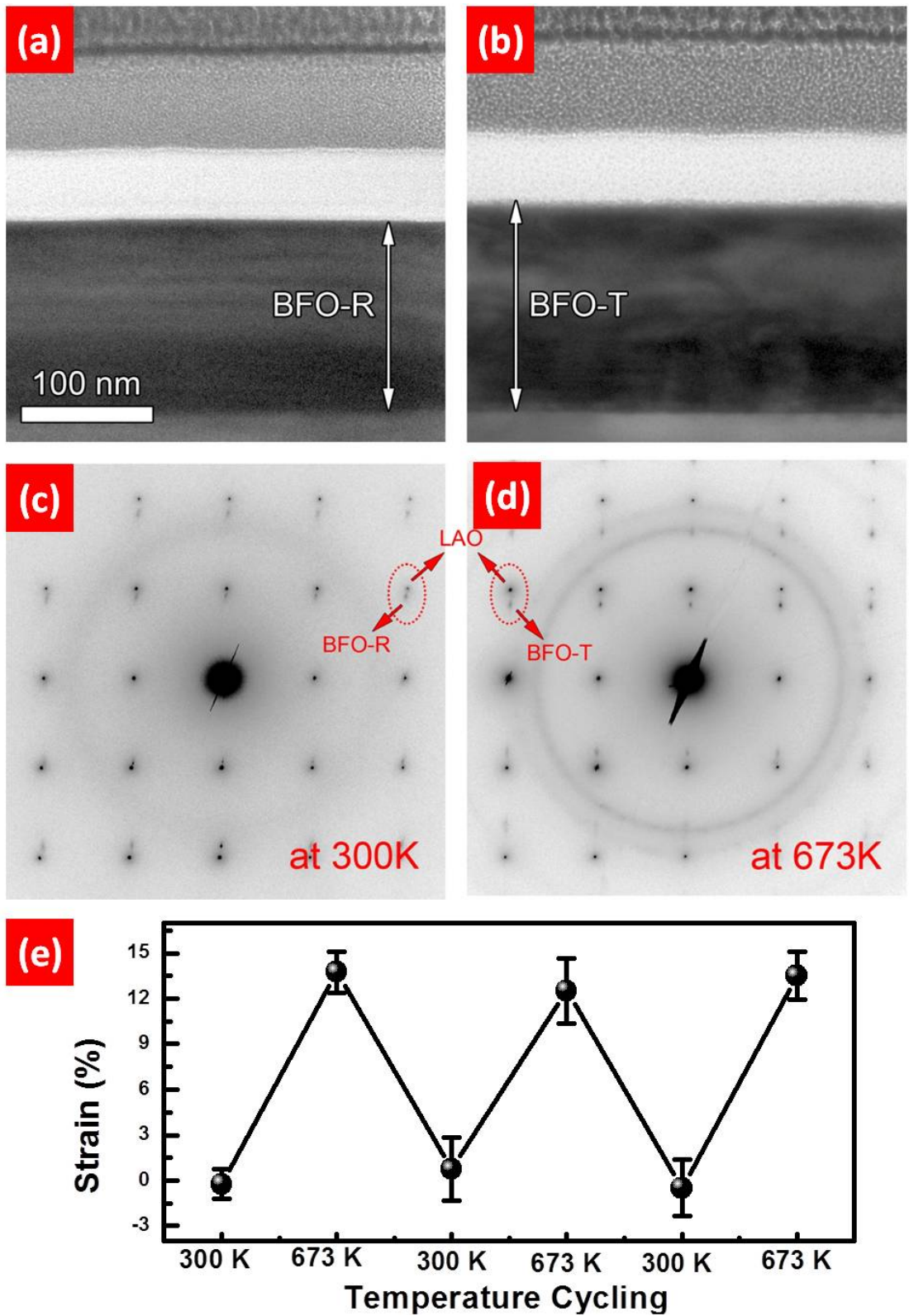

**Fig. 4. Zhang et al.**

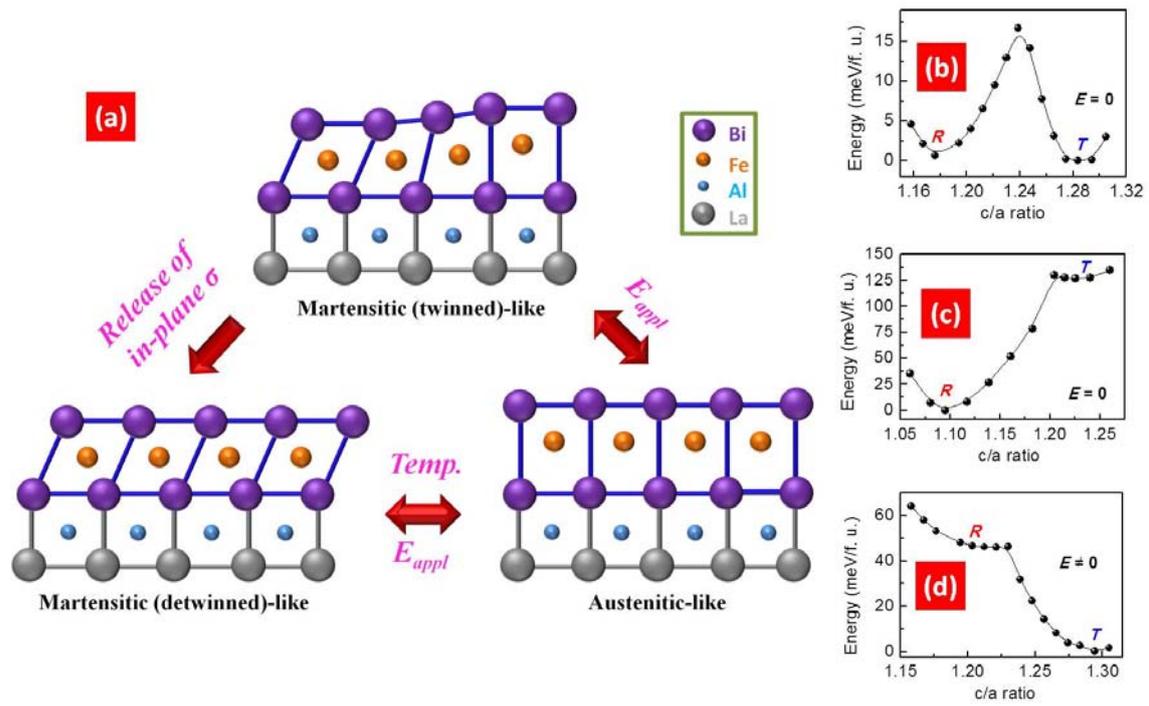

Fig. 5. Zhang et al.